\documentclass[12pt,prd,aps,amssymb,amsmath,tightenlines,showpacs,a4paper]{revtex4}
\def\half{\textstyle{\frac{1}{2}}}
\def\sixth{\textstyle{\frac{1}{6}}}
\def\cP{{\cal P}}
\def\cC{{\cal C}}
\def\cT{{\cal T}}
\def\cQ{{\cal Q}}
\begin{document}

\title{Comment on ``New ansatz for metric operator calculation in pseudo-Hermitian field theory''}

\author{Carl~M.~Bender${}^1$}\email{cmb@wustl.edu}
\author{Gregorio Benincasa${}^2$}\email{gregorio.benincasa08@imperial.ac.uk}
\author{H.~F.~Jones${}^2$}\email{h.f.jones@imperial.ac.uk}

\affiliation{${}^1$Department of Physics, Washington University, St. Louis, MO
63130, USA \\
${}^2$Physics Department, Imperial College, London SW7 2BZ, UK}


\begin{abstract}
In a recent Brief Report by Shalaby a new first-order perturbative calculation
of the metric operator for an $i\phi^3$ scalar field theory is given. It is
claimed that the result is an improvement on
a previous calculation by Bender, Brody and Jones because it is local. Unfortunately
Shalaby's calculation is not valid because of sign errors.
\end{abstract}

\maketitle

The possibility that a non-Dirac-Hermitian $\cP\cT$-symmetric Hamiltonian, such
as $H=p^2+ix^3$, could define a consistent quantum-mechanical theory was first
pointed out in Ref.~\cite{r1}. It was subsequently shown \cite{r2,r3,r4} that
such Hamiltonians are self-adjoint with respect to the $\cC\cP\cT$ operator,
where $\cC$ is a linear operator satisfying the following system of three
simultaneous algebraic equations:
\begin{equation}
\cC^2=1,\qquad [\cC,\cP\cT]=0,\qquad [\cC,H]=0.
\label{e1}
\end{equation}

For cubic Hamiltonians of the general form $H=H_0+\epsilon H_1$,
where in quantum mechanics
\begin{equation}
H=\half p^2+\half x^2+i\epsilon x^3
\label{e2}
\end{equation}
and in quantum field theory
\begin{equation}
H=\int d^Dx\left[\half\pi^2+\half(\nabla\phi)^2+\half m^2\phi^2+i\epsilon\phi^3
\right],
\label{e3}
\end{equation}
it is explained in Refs.~\cite{r3,r4} how to calculate the $\cC$ operator using
perturbative methods. The procedure is to take the $\cC$ operator to have the
form
\begin{equation}
\cC=e^\cQ\cP,
\label{e4}
\end{equation}
where $\cQ$ is a Hermitian operator that has a formal power series in odd powers
of $\epsilon$: $Q=\epsilon\cQ_1+\epsilon^3\cQ_3+\epsilon^5\cQ_5+\cdots$.
Substituting (\ref{e4}) into (\ref{e1}) and collecting powers of $\epsilon$, one
obtains a sequence of algebraic equations that can be solved successively to
determine the coefficients $\cQ_1$, $\cQ_3$, $\cQ_5$, and so on. The first three
of these equations are
\begin{eqnarray}
[\cQ_1,H_0]&=&2H_1,\cr
[\cQ_3,H_0]&=&\sixth[\cQ_1,[\cQ_1,H_1]],\cr
[\cQ_5,H_0]&=&-\textstyle{\frac{1}{360}}[\cQ_1,[\cQ_1,[\cQ_1,[\cQ_1,H_1]]]]
+\sixth[\cQ_1,[\cQ_3,H_1]]+\sixth[\cQ_3,[\cQ_1, H_1]].
\label{e5}
\end{eqnarray}

In Ref.~\cite{r4}, Bender, Brody, and Jones calculated $\cQ_1$ for a cubic
quantum field theory in $D+1$-dimensional space-time defined by (\ref{e3}). They
found an expression for $\cQ_1$ as a nonlocal function of the field operator
$\phi$. We emphasize that the result for $\cQ_1$ in Ref.~\cite{r4} is correct.

Recently, Shalaby reported a new calculation of $\cQ_1$ in Ref.~\cite{r5}, in which he
obtained a different result from that in given in Ref.~\cite{r4}. It is not surprising
that there might be another solution for $\cQ_1$ because, as is shown in
Ref.~\cite{r6}, the solution to the algebraic equations (\ref{e1}) is not
unique. Shalaby's reported solution is claimed to be {\it local} and this new solution
is characterized as being less ``cumbersome'' than the earlier nonlocal result of Ref.~\cite{r4}.

Unfortunately, the calculation reported by Shalaby in Ref.~\cite{r5} is wrong.
The first error may be found in the fifth unnumbered
equation after Eq.~(4). This equation is said to contain a total derivative that
integrates to zero, but the expression is only a total derivative when the
space-time dimension is 2. The correct form of the integrand is in fact
$\nabla^2\phi(x)[\nabla\phi(x)]^2$, which in general is not a total derivative.
The second error, which occurs in the fifth line of the unnumbered equation before
Eq.~(3), is an incorrect treatment of the derivative of a delta function. This leads
to incorrect signs in two further equations. When the signs are corrected, the resulting
equations for the coefficients $C_1$, $C_2$, and $C_3$ become inconsistent, thus
invalidating the reported result.

\begin{acknowledgments}
CMB is grateful to the Theoretical Physics Group at Imperial College for its
hospitality and the U.S.~Department of Energy for financial support.
\end{acknowledgments}

\end{document}